\documentclass[prb,aps,amssymb,amsmath,reprint,showpacs,floatfix,superscriptaddress,longbibliography]{revtex4-2}
\usepackage[utf8]{inputenc}
\usepackage{graphicx}
\usepackage{color}
\usepackage{amsmath}
\usepackage{dsfont}
\usepackage{amsfonts}
\usepackage{amssymb}
\usepackage{hyperref}
\usepackage{braket}
\usepackage{appendix}

\newcommand{\rhoi}{\rho^{}_{\mathrm{i}}}
\newcommand{\rhof}{\rho^{}_{\mathrm{f}}}
\newcommand{\Po}[1]{P^{}_{#1}}

\hypersetup{colorlinks=true}
\hypersetup{citecolor=cyan}

\begin{document}

\title{Quantum process tomography of a compressed time evolution circuit on superconducting quantum processors}

\author{Maria Dinc\u a}
\affiliation{Department of Materials, University of Oxford, 12-13 Parks Road, Oxford OX1 3PU, United Kingdom}
\author{David J. Luitz}
\affiliation{Institute of Physics, University of Bonn, Nu\ss allee 12, 53115 
Bonn, Germany}
\author{Maxime Debertolis}
\affiliation{Institute of Physics, University of Bonn, Nu\ss allee 12, 53115 
Bonn, Germany}

\begin{abstract} 
    
As present day quantum hardware is limited by various noise mechanisms, quantum advantage can only be reached in the near-term by designing noise-resilient quantum algorithms.
In this work, we employ state-of-the-art quantum process tomography (QPT) techniques to characterize the noise channels of IBM quantum processors under realistic runtime constraints. As our main application, we compare the Trotter time-evolution of three- and four-qubit wave functions to a compressed quantum circuit version of the same evolution operator. By analysing the spectral properties of the two process channels, we find that the compressed circuit systematically yields larger eigenvalue moduli, demonstrating better noise resilience. 
\end{abstract}

\maketitle

\section{Introduction}
\label{Introduction}
	Quantum processors are progressing at a remarkable pace and are widely anticipated to eventually surpass the computational power of classical computers~\cite{Feynman_1982,Lloyd_1996,Shor1997,Huang2022,king2024}. While the long-term goal of a fully error-corrected, fault-tolerant quantum computer remains out of reach with current technology, Noisy Intermediate-Scale Quantum (NISQ) devices provide an essential platform for exploring near-term quantum applications~\cite{Preskill_2018,Boixo_2018,Bharti_2022}. These intermediate devices have proven over the years to be capable of performing meaningful experiments in many fields of physics~\cite{Aspuru-Guzik_2005,Lanyon2011,Heras2014,Salath_2015,Acín_2018,Perdomo-Ortiz_2018, Zhong_2020, GoogleAI2020, Satzinger2021,Funcke2021,Richter_2021}, despite their inherent noise and limited qubit numbers.

Among the variety of experimental platforms under active development, trapped 
ions~\cite{Haffner_2008,Ballance_2016}, superconducting 
circuits~\cite{Barends_2014,Huang_2020_review} and neutral atom platforms 
exploiting atomic Rydberg states~\cite{briegel_quantum_2000, 
bluvstein_quantum_2022} have emerged as the most promising candidates. 
Trapped-ion qubits offer long coherence times and high-fidelity entangling 
operations, but are limited by relatively slow gate times and difficulties in 
scaling up the architecture. Neutral atom platforms face similar scaling 
challenges but bring the advantage of free reconfiguration of qubits to 
implement arbitrary connectivity~\cite{bluvstein_quantum_2022}. Silicon spin 
qubits can be integrated with existing CMOS technology, allowing for scalability 
via reduced physical sizes, but achieving flexible connectivity is still a challenging task~\cite{siegel2025snakesplanemobilelow}.
In contrast, superconducting platforms work with a fixed processor topology, but 
benefit from fast gate operations, and ongoing progress in scaling to larger 
qubit arrays. This makes them attractive for exploring algorithms and simulation 
protocols.

One of the most promising near-term applications of quantum hardware is the 
simulation of quantum many-body systems~\cite{Zalka_1998,Fauseweh_2024}, where 
the exponential complexity of classical computations is naturally avoided on 
quantum processors. However, current superconducting devices suffer from a 
number of limitations: qubit connectivity is typically restricted to nearest 
neighbors on some kind of two dimensional lattice, and every quantum gate 
introduces errors due to the required coupling with the environment. As more 
gates are applied, these errors accumulate and eventually dominate the 
computation, rendering the output indistinguishable from random noise.

Several strategies have been developed to deal with this intrinsic vulnerability. A major direction is quantum error mitigation~\cite{Temme_2017, Cai_2023}, which encompasses protocols to estimate or suppress noise without requiring the space-time overhead of full quantum error correction. These approaches include 
extrapolation techniques~\cite{Li_2017,Cai_2021} or probabilistic error 
cancellation~\cite{McDonough_2022,Takagi_Endo_Minagawa_Gu_2022,van_den_Berg_2023,Gupta2024}, all of which aim to average out errors by performing additional measurements. Combining digital quantum simulation with error-mitigation techniques has already yielded encouraging results, allowing the reproduction of physically relevant processes such as the real-time evolution of interacting quantum systems~\cite{Kim_Eddins_2023,Fauseweh_2024}. Although powerful, these methods come with significant measurement overheads. 

A complementary approach to reducing noise is to directly shorten the depth of 
quantum circuits, thereby limiting the number of noisy operations. This has 
motivated the development of compressed 
circuits~\cite{Mansuroglu_Eckstein_2023,Mansuroglu_2023,McKeever2023,Tepaske_2023,Tepaske_2024}, 
which provide optimized approximations of a given process—such as Hamiltonian 
time evolution—using parameterized gate sequences of tunable depth and 
connectivity. Circuit compression algorithms, typically designed and executed on 
classical computers, identify optimal representations that can reduce the depth 
by nearly 50\% compared to standard Trotter–Suzuki decompositions at fixed 
fidelity~\cite{Tepaske_2023}. In principle, such reductions are especially 
valuable for superconducting platforms, where non-local interactions require 
additional SWAP gates, which are particularly noisy and can lead to significant error propagation. So far, 
compressed circuits have been mostly studied in theory or numerical simulations, 
under simplified noise models. Their practical 
performance under realistic hardware noise remains largely untested. In this 
article, we address this open question by confronting compressed circuits with 
standard Trotter decompositions in experimental realizations on quantum 
hardware.

    To quantitatively assess the quality of the implemented processes, we make 
use of quantum process tomography (QPT)~\cite{Chuang_1997, 
OBrien2004,Howard_2006,Mohseni_Rezakhani_Lidar_2008,Nielsen_Chuang_2010}. QPT 
reconstructs the quantum channel describing the implemented process, without 
relying on assumptions about input states or device-specific noise, thus 
providing an unbiased characterization of the experiment. From the reconstructed 
process matrix, one can not only compute fidelities with respect to target 
processes but also identify sources of decoherence or systematic errors in the 
implementation~\cite{Kofman_Korotkov_2009,Kimmel_daSilva_Ryan_Johnson_Ohki_2014,Flammia_Wallman_2020,van_den_Berg_2023,Onorati_2023}. 
However, QPT scales poorly with system size, as the number of parameters describing the channel grows 
exponentially with the number of qubits.

	For this reason, our study focuses on small systems, where QPT is still feasible, but already reveals clear differences between circuit implementations. Specifically, we demonstrate the use of compressed circuits on three- and four-qubit Heisenberg spin chains. For the three-qubit case, we perform a full process tomography to characterize the implemented processes in detail. For four qubits, where the full-QPT becomes computationally prohibitive, we employ Pauli twirling~\cite{Wallman_2016, knill2004} and selective quantum process tomography (SQPT)~\cite{Bendersky_Pastawski_Paz_2008, Bendersky_Pastawski_Paz_2009,Schmiegelow_Bendersky_Larotonda_Paz_2011,Perito_Roncaglia_Bendersky_2018,Stefano_Perito_Rebon_2023}, which allows the efficient reconstruction of relevant elements of the process matrix. This combination enables us to benchmark compressed circuits against Trotter decompositions under realistic noise conditions and to demonstrate the advantage of circuit compression in experimental quantum simulations. To the best of our knowledge, this is the first experimental implementation of SQPT beyond three qubits and for deep-circuit channels rather than individual gates, as well as the first use of Pauli twirling for tomography purposes.

\section{Time evolution on a quantum computer}

	\begin{figure}[t]
                {\centering \includegraphics[width=0.45\columnwidth]{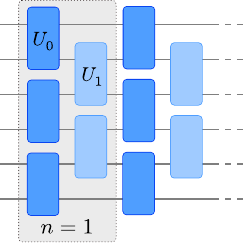}} 
		\caption{Quantum circuit of the $1^{\mathrm{st}}$ order Trotter decomposition of the nearest-neighbor Heisenberg Hamiltonian with open boundary conditions.  The qubits are arranged on a linear grid, consistent with the topology of the device on which the experiments are performed. The gray frame highlights a single Trotter step alternating nearest-neighbor terms acting on even and odd bonds, which is repeated to achieve the desired time evolution.}
		\label{fig::Trott1}
	\end{figure}	

	For concreteness, we study the time evolution operator of the the $SU(2)$ symmetric Heisenberg model of $s=1/2$ spins on a lattice with $L$ qubits with neareast neighbor couplings, whose Hamiltonian reads:
	\begin{equation}
		\label{eq::heis}
		\mathcal{H}^{}_{\mathrm{heis}} = \sum\limits^{L}_{\langle i,j \rangle} \vec{S}_{i} \cdot \vec{S}_{j}.
	\end{equation}
We consider open and periodic boundary conditions for chain length of $L=3$ and 
$L=4$ respectively, for which the process tomography can be performed. While the 
connectivity of the Hamiltonian is dictated by the system of interest, it does 
not necessarily match the connectivity of the quantum processor. We impose periodic 
boundary conditions for the system with $L=4$ qubits to highlight this mismatch. 
Even though in this simple case it would be possible to arrange qubits in a 
ring, we deliberately use a different topology on the processor to investigate 
how circuit compression can help to avoid extraneous SWAP 
gates~\cite{Tepaske_2023}.
We introduce two representations of the time evolution operator as a quantum 
circuit that we confront in this study: Trotter and compressed circuits. \\

	\emph{Trotter-Suzuki decomposition:}
		\begin{figure}[t]
                        \centering
                        \includegraphics[width=1\columnwidth]{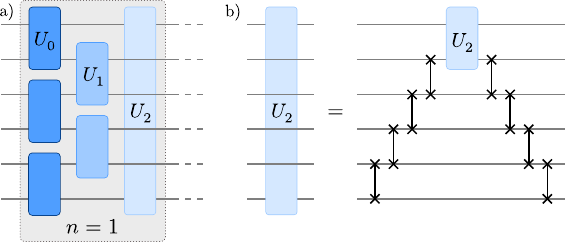}
			\caption{(a) Quantum circuit of the $1^{\mathrm{st}}$ order Trotter decomposition with periodic boundary conditions, in which the gate $U^{}_{2}$ couples the first and last qubit. (b) On a device implementing only nearest neighbor gates on a linear topology, a set of swap gates has to be implemented to connect the two distant qubits.}
			\label{fig::Trott2}
		\end{figure}
The standard approach to simulate the time evolution of a Hamiltonian such as 
that in Eq.~\eqref{eq::heis} is the Trotter-Suzuki (or simply Trotter) 
decomposition~\cite{Suzuki_1990}, which approximates the unitary operator governing 
the dynamics. This method exploits the Baker–Campbell–Hausdorff formula, which 
expresses the exponential of a sum of non-commuting operators as a series of 
nested commutators. Truncating this series introduces an error proportional to 
the neglected terms, which can be reduced by subdividing the total evolution 
time $t$ into $n$ successive steps of duration $dt = t/n$. At first order, the 
exact propagator $U^{}_{\textsc{E}}$ can be approximated by:
		\begin{equation}
            U^{}_{\textsc{e}} = e^{-i\hat{\mathcal{H}}t}_{} 
            \simeq U^{\mathrm{(1)}}_{\textsc{t}}(n) = \left( 
                \prod\limits^{N^{}_{\textsc{h}}}_{\alpha=1}
            e^{-i\hat{H}_{\alpha}dt}_{} \right)^{n},
		\end{equation}
where the Hamiltonian $\hat{\mathcal{H}}$ is partitioned into $N^{}_{\textsc{h}}$ commuting components $\hat{H}^{}_{\alpha}$, each containing $N^{}_{\alpha}$ two-qubit operators in the present model:
		\begin{equation}
			\hat{\mathcal{H}} = \sum\limits^{N^{}_{\textsc{h}}}_{\alpha=1} \hat{H}^{}_{\alpha}, \quad \text{with} \quad \hat{H}^{}_{\alpha} = \sum\limits^{N^{}_{\alpha}}_{\beta=1}
			\hat{h}^{\beta}_{\alpha}.
		\end{equation}
In our case, the Hamiltonian is split into $N^{}_{\textsc{h}}=2$ subsets 
consisting of nearest-neighbor terms acting on even and odd bonds, respectively. 
Since each factor in the decomposition is unitary, the approximate propagator 
remains unitary. Figure~\ref{fig::Trott1} illustrates the circuit representation 
of this Trotter evolution with open boundary conditions, where $U^{}_{0}$ 
and $U^{}_{1}$ correspond to the even- and odd-bond contributions. The shaded 
block in the figure implements evolution over $dt$, and its repetition $n$ times 
yields evolution up to $T = n dt$.

On the quantum devices used in this work, each gate $U^{}_{\alpha}$ must be further decomposed 
into elementary operations, namely one-qubit rotations and two-qubit {\sc cx} 
gates. Specifically, a Trotter gate acting on neighboring qubits takes the form 
$U^{}_{\alpha} = \mathrm{exp}\left(-i dt \sum_{k} \sigma^{k}_{i}\sigma^{k}_{i+1} 
\right)$, which can be realized with 3 two-qubit {\sc cx} gates and 5 
single-qubit {\sc rx} gates, as shown in the circuit shown in Fig.~\ref{fig:trotter_circ}:

        \begin{figure}[h!]
                        {\centering 
        \includegraphics[width=0.7\columnwidth]{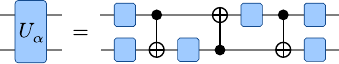}}
        \caption{Decomposition of a general unitary gate into single qubit and 
            CNOT gates. \label{fig:trotter_circ}}
		\end{figure}

		For periodic boundary conditions, the Hamiltonian can no longer be decomposed into only two sets of nearest-neighbor terms, and an additional non-local gate is required to couple the first and last qubits, as depicted in Fig.~\ref{fig::Trott2}.~a). Since the quantum hardware under consideration features only nearest-neighbor connectivity, implementing this long-range gate entails an overhead of $\mathcal{O}(L)$ swap operations, which transfer the logical states of the distant qubits on adjacent qubits, so that local gates can be applied, as illustrated in Fig.~\ref{fig::Trott2}.~b). Consequently, the circuits simulating periodic boundary conditions are deeper and therefore more susceptible to noise.

		As discussed above, first-order Trotter circuits approximate the exact time evolution operator $U_{\textsc{e}}$ with an approximation error scaling as $\mathcal{O}(dt^{2})$. Higher-order decompositions improve the accuracy but require additional operations. In our case, where $N_{\textsc{h}}$ is close to one, a second-order scheme introduces nearly the same number of operations as the first-order one, while reducing the approximation error to $\mathcal{O}(dt^{3})$:
		\begin{equation}
			U^{\mathrm{(2)}}_{\textsc{t}} = \left( \prod\limits^{1}_{\alpha=N_{\textsc{h}}} e^{-i\hat{H}_{\alpha}\frac{dt}{2}}_{}
			\prod\limits^{N_{\textsc{h}}}_{\alpha=1} e^{-i\hat{H}_{\alpha}\frac{dt}{2}}_{}\right)^{n}_{}.
		\end{equation}
		With periodic boundary conditions, the non-local gate is inserted at the midpoint of each Trotter step, allowing the two half-step contributions ($dt/2$) to be combined into a single $dt$ contribution, thereby reducing the number of extra gates:
		\begin{equation}
			U^{\mathrm{(2)}}_{\textsc{t}}\hspace{-0.05cm}= \hspace{-0.05cm}\Big( U^{}_{0}\hspace{-0.1cm}\left(\frac{dt}{2}\right)\hspace{-0.05cm} U^{}_{1} \hspace{-0.1cm}\left(\frac{dt}{2}\right)\hspace{-0.05cm} U^{}_{2}(dt) U^{}_{1}\hspace{-0.1cm}\left(\frac{dt}{2}\right)\hspace{-0.05cm} U^{}_{0}\hspace{-0.1cm}\left(\frac{dt}{2}\right)\hspace{-0.05cm} \Big)^{n}_{} \hspace{-0.13cm}.
		\end{equation}
		Similarly, consecutive $U^{}_{0}$ operations for $n>1$ can be merged into a single gate with duration $dt$. In this work, we do not pursue higher-order decompositions, since the additional noisy gates required at each layer prevent reliable results. \\

	\emph{Compressed circuit:}
		An efficient realization of time evolution on a quantum processor can be achieved by encoding the dynamics into a compressed circuit. Such a circuit corresponds to an optimized unitary operator obtained from a variational ansatz, designed both to approximate the exact unitary evolution and to match the constraints of the target hardware. The parametrized unitary operator is expressed as a product of a fixed number $N_{\textsc{g}}$ of gates:
		\begin{equation}
			\label{eq::compressedC}
			U^{}_{\textsc{c}}\big(\,\vec{\theta}\,\big) = \prod\limits^{N^{}_{\textsc{g}}}_{\alpha=1} \mathrm{mat}\left(
			U^{}_{i^{}_{\alpha}j^{}_{\alpha}}\big(\,\vec{\theta}^{}_{i^{}_{\alpha}}\big)  \right),
		\end{equation}
		where $\mathrm{mat(U)}$ denotes the $(2^{L}\times2^{L})$ unitary representation of the $\alpha^{\mathrm{th}}$ gate in the circuit. For an ansatz composed of nearest-neighbor two-qubit gates, each contribution in Eq.~\eqref{eq::compressedC} takes the form
		\begin{equation}
			\mathrm{mat} \left(U^{}_{ij} \right) = \mathds{1}_{2^{i-1}_{}} \otimes U^{}_{ij} \otimes \mathds{1}_{2^{L-j}_{}}  
		\end{equation}

\begin{figure}[t]
                   \centering
                   \includegraphics[width=0.8\columnwidth]{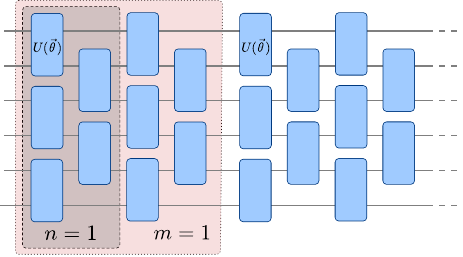} 
			\caption{Compressed quantum circuit constructed in a brickwork structure. Each gate in the approximated shallow circuit corresponding to the $m=1$ shaded area is parametrized by a 12-dimensional vector $\vec{\theta}$ (see text) in order to approximate $U(t)$. The compressed circuit can be repeated to reach later times, which are multiples of the approximated time evolution operator, $U(mt)$. The depth of the compressed circuit is counted by its number of brickwall layers $n$.}
			\label{fig::Compressed_circs}
		\end{figure}

		Unlike the Trotter decomposition, which is tied to the Hamiltonian structure and often produces circuits whose topology does not match with the hardware connectivity—leading, for instance, to additional swap operations under periodic boundary conditions—the compressed circuit approach is flexible and can be tailored to the device. This adaptability enables a significant reduction in noisy operations. To maximize the accessible unitary space while respecting the linear connectivity of the qubits, we adopt a brickwall circuit layout for the sequence of gates in Eq.~\eqref{eq::compressedC}, as illustrated in Fig.~\ref{fig::Compressed_circs}.

		Beyond the overall architecture, each two-qubit gate 
        $U_{ij}(\vec{\theta})$ must be parametrized. Requiring only unitarity, 
        such a gate can be decomposed into single- and two-qubit 
        components~\cite{Vatan2004}:
		\begin{equation}
			U^{}_{ij} = \left(U^{}_{i} \otimes U^{}_{j} \right) \, V^{}_{ij} \, \left(U^{'}_{i} \otimes U^{'}_{j} \right),
		\end{equation}
        which corresponds to the circuit \begin{center} 
 \includegraphics[width=0.27\textwidth]{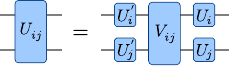}
\end{center}

        Here, each single-qubit operation $U$ or $U^{'}$ depends on three 
        parameters:
		\begin{equation}
			U^{}_{i}\left(\mu^{}_{1}, \mu^{}_{2}, \mu^{}_{3}\right) = 
			\begin{pmatrix}
				e^{i\mu^{}_{2}}\cos\left(\mu^{}_{1} \right) & e^{i\mu^{}_{3}}\sin\left(\mu^{}_{1} \right) \\
				-e^{-i\mu^{}_{3}}\sin\left(\mu^{}_{1} \right) & e^{-i\mu^{}_{2}}\cos\left(\mu^{}_{1} \right)
			\end{pmatrix}.
		\end{equation}
		while the two-qubit gate $V_{ij}$ acts on adjacent qubits and is parametrized by three additional parameters:
		\begin{equation}
			V^{}_{ij}\left(\nu^{}_{1}, \nu^{}_{2}, \nu^{}_{3} \right) = \mathrm{exp}\Big(-i \sum\limits^{3}_{k=1} \nu^{}_{k}\,\sigma^{k}_{i}\otimes\sigma^{k}_{j}\Big).
		\end{equation}
		The single-qubit rotations $U$ and $U^{'}$ are natively supported by hardware, while $V_{ij}$ is implemented through the same two-qubit decomposition of the gates $U_{\alpha}$ building Trotter circuits. In practice, consecutive single-qubit gates between two layers are merged to minimize depth, yielding 12 parameters per gate in the bulk, with the final layer consisting only of single-qubit rotations.

        The optimal parameters $\vec{\theta}$ are determined classically using 
        the gradient-based optimizer {\sc{adam}}~\cite{KingmaBa2017}. The cost 
        function minimized during the training is the approximation error (using the same terminology as for the Trotterization procedure), 
        defined as the Frobenius norm distance between the exact evolution 
        operator $U_{\textsc{e}}$ and its approximation as a compressed circuit 
        $U_{\textsc{c}}$:
		\begin{equation}
			\label{eq::infidelity}
			\epsilon = \frac{1}{2} ||U^{}_{\textsc{e}} - U^{}_{\textsc{c}}||^{2}_{\textsc{f}} = 1 - \frac{\mathrm{Re}\left( \mathrm{Tr}\left[ U^{\dagger}_{\textsc{e}}U^{}_{\textsc{c}}\right]\right)}{2^{L}}.
		\end{equation}
		For the small system sizes considered in this work, $U_{\textsc{e}}$ can be obtained exactly in the full Hilbert space, leading to smooth and fast convergence. The approximation error of the compressed circuit depends on the number of variational parameters, which increases with the number of brickwall layers that are considered.

		\begin{figure}[t]
                   \centering
                   \includegraphics[width=1\columnwidth]{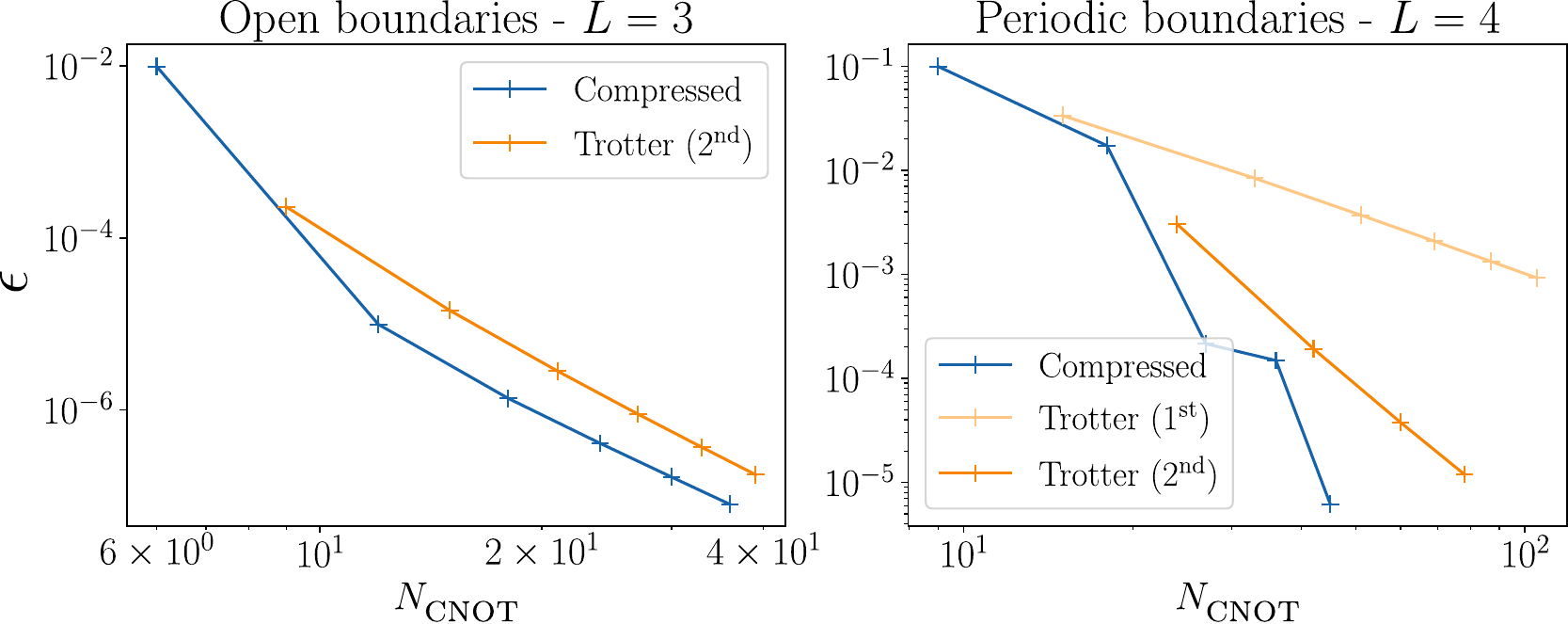}
			\caption{\textit{Left:} Approximation error $\epsilon$ as written in Eq.~\eqref{eq::infidelity} for the $2^{\mathrm{nd}}$ order Trotter and compressed approximations of the exact unitary evolution at time $t=1$. The number of {\sc{cnot}}s is estimated from the individual decomposition of each two-qubit gate in the considered circuit, and no further optimization than presented in the main text is considered at this level. \textit{Right:} Approximation error of the $1^{\mathrm{st}}$ and $2^{\mathrm{nd}}$ order Trotter decompositions, and of the compressed circuit in presence of periodic boundary conditions.} 
			\label{fig::Infid}
		\end{figure} 

		To compare compressed and Trotter circuits, we use the approximation error $\epsilon$ of Eq.~\eqref{eq::infidelity} as a convergence criterion for a fixed number of {\sc{cnot}} gates, the dominant source of noise. Results for a three-qubit nearest-neighbor Heisenberg model and for a four-qubit system with periodic boundary conditions are shown in Fig.~\ref{fig::Infid}. In order to tune the approximation error of the Trotter circuit, we change the number of layers $n$ achieving the time evolution up to time $t=1$, hence changing the time steps $dt$. For the compressed circuit, we simply tune the number of layers $n$, allowing more coefficients in the circuit ansatz. For open boundaries, the second-order Trotter scheme already achieves good performance because the circuit layout matches the device connectivity, reducing the advantage of compression. For periodic systems, however, the Trotter gate count grows rapidly, such that only the first-order scheme remains practical, making the compressed circuit a more favorable alternative.
		In the following section, we discuss process tomography methods employed to characterize noise in both circuit types, focusing on the situations where they have a comparable approximation error $\epsilon$.

\section{Quantum Process Tomography}
	\label{Full QPT}
	
The evolution of a quantum state under a physical process can be formally 
described by a completely positive map that connects the initial and final
density matrices ($\rhoi$ and $\rhof$, respectively). This map admits the Kraus representation:
	\begin{equation}
		\label{KrausMap}
		\Lambda(\rhoi) = \sum\limits^{}_{a}\mathcal{K}^{}_{a} \, \rhoi \mathcal{K}^{\dagger}_{a} = \rhof.
	\end{equation} 
		Although the Kraus representation is not unique, the set of operators ${\mathcal{K}_{a}}$ fully characterizes the quantum process superoperator $\Lambda$. In the following, we adopt the notation $L$ for the number of qubits, $D=2^{L}$ for the Hilbert space dimension, and $N=4^{L}$ for the number of elements in any operator basis on this space.

	\begin{figure*}[t!]
		\centering
		\includegraphics[width=0.8\textwidth]{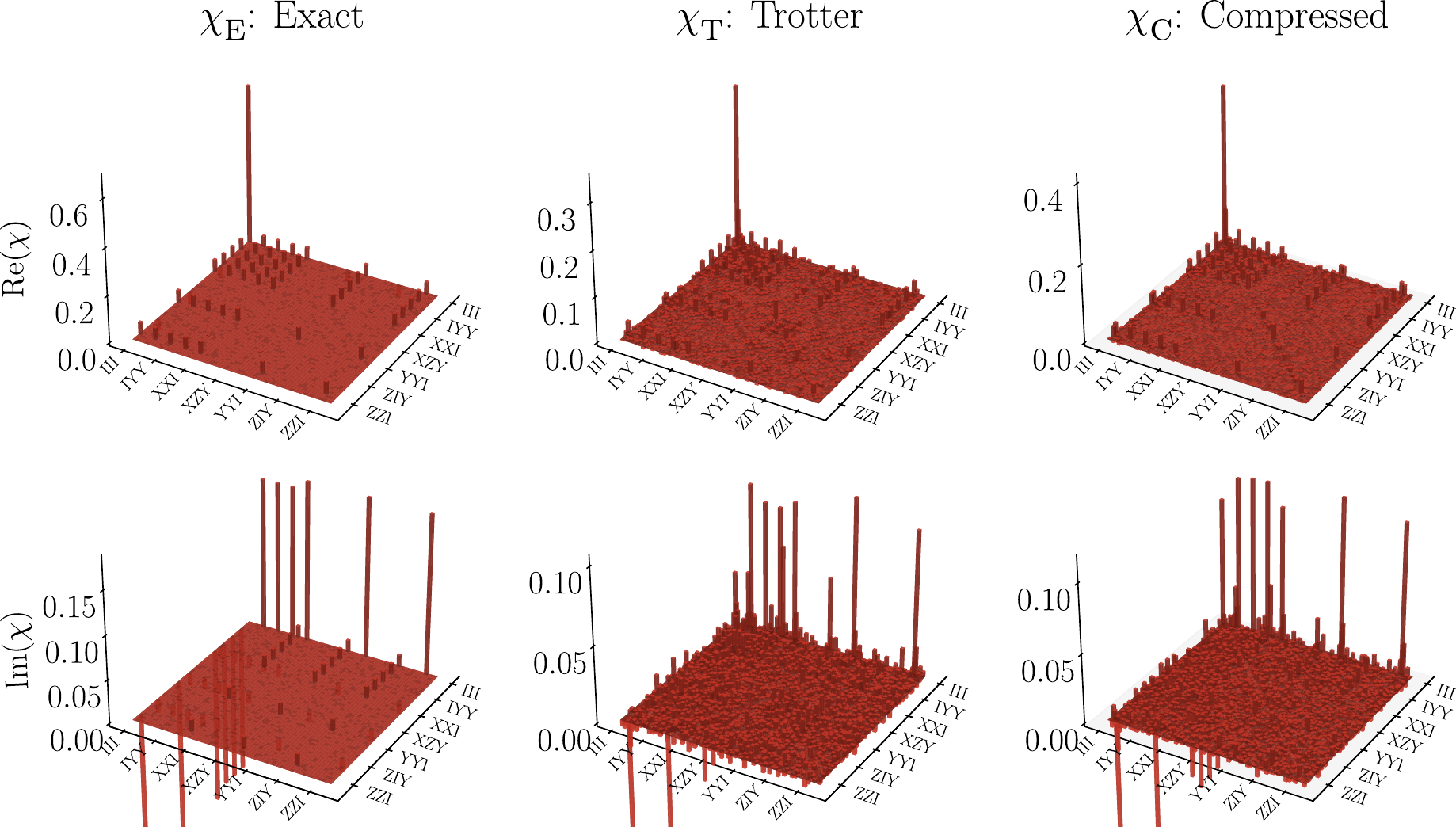}
		\caption{Real and imaginary parts of the $\chi$ matrices for the ideal circuit (left panels) representing the time evolution operator up to time $t=1$ of the Heisenberg model with $L=3$ and open boundary conditions. The middle and right panels show the experimentally reconstructed process matrices for the Trotter circuit and the compressed circuit respectively. The Trotter circuit consists of a second order Trotter decomposition with two layers such that $dt=0.5$, yielding to an approximation error of $\epsilon=1.5\times 10^{-5}$, and the fidelity of the process is $F_{\textsc{T}}=48\%$. The compressed circuit also has two layers of gates in a brickwall, leading to $\epsilon = 1.0\times 10^{-5}$ and $F_{\textsc{C}}=64\%$.}
		\label{fig::Chi}
	\end{figure*} 

		To make Eq.~\eqref{KrausMap} amenable to computation, Kraus operators are expanded in a basis of $D$-dimensional matrices, typically chosen to be the Pauli basis. This basis consists of $N$ Pauli strings, defined as $L$-fold tensor products of single-qubit Pauli operators (including the identity): $P_{i} \in \{\mathds{1}, X, Y, Z\}^{\otimes L}$. In this representation, the map $\Lambda$ is expressed through the $\chi$ matrix:
	\begin{equation}
		\label{chimat}
		\Lambda(\rhoi) = \sum\limits^{N}_{m,n=1} \chi^{}_{mn} \Po{m} \, \rhoi  \Po{n}, \quad \chi^{}_{mn} = \sum\limits^{}_{a} b^{}_{am} b^{*}_{an},
	\end{equation}
	where the coefficients $b_{am}$ correspond to the weight of each $\mathcal{K}_{a}$ in the Pauli basis. Owing to the Hermitian property of Pauli strings ($P^{\dagger}_{m} = P_{m}$), the process matrix $\chi$ has the following properties: it is also Hermitian ($\chi=\chi^{\dagger}$), positive semidefinite (all eigenvalues $\lambda_{n}\geq 0$), and non–trace-increasing ($\sum^{N}_{m,n=1} \chi_{mn} P_{m} P_{n} \leq {\mathds{1}}_{D}$). The $\chi$ matrix thus encodes the full information about the quantum process, and can be reconstructed from appropriate input state preparations and measurements. To this end, the output state is first decomposed in the Pauli basis:
	\begin{equation}
		\label{measure_tomo}
		\rhof = \sum^{N}_{k} d^{}_{fk} \Po{k},
	\end{equation}
	with $d_{fk} = \langle \Po{k} \rangle_f = \mathrm{Tr}[\rhof \Po{k}] / \mathrm{Tr}[\rhof]$, corresponding to the experimentally accessible expectation value of $\Po{k}$ in the final state. Similarly, the input state can be expanded as $\rhoi=\sum_l c_{il}\Po{l}$. Since products of Pauli strings are proportional to Pauli strings ($P_m P_l P_n \propto P_k$), the coefficients in Eq.~\eqref{chimat} and Eq.~\eqref{measure_tomo} can be related, leading to the following linear system for $\chi$:
	\begin{equation}
		\kappa \vec{\chi} = \vec{d}, \quad \text{with} \quad \kappa^{}_{(mn),(kl)} \Po{k} = \Po{m}\Po{l}\Po{n},
	\end{equation}
	where the $D$-dimensional operators have been reshaped into $N$-dimensional vectors. Solving this system requires inversion of the matrix $\kappa$ with dimension $4^{2L}$, which is classically tractable only for small systems ($L \leq 3$). Once $\chi$ is obtained, it can be diagonalized to extract the corresponding Kraus operators of Eq.~\eqref{KrausMap}.

	While the characterization of the full process is in general desirable, the 
    measurement overhead grows exponentially with $L$, preventing us from 
    addressing periodic boundary conditions that require at least four qubits. 
    Moreover, the Hamiltonian evolution considered here exhibits significant structure in the 
    Pauli basis, leading to highly sparse $\chi$ matrices. Exploiting this 
    sparsity is essential to reduce both experimental and computational costs. 
    One strategy to achieve such efficient reconstruction for the four-qubit 
    case is selective quantum process tomography (SQPT), which we explain in 
    Appendix~\ref{app:sqpt}. However, SQPT introduces a fixed overhead of 
    additional gates applied on top of the target process. For the small 
    circuits considered here, this overhead is comparable to the number of gates 
    implementing the time evolution itself. We therefore focus in the following 
    primarily on the three-qubit case, and defer the discussion of four-qubit 
    results to the end.

\section{Results}

	\begin{figure}[t]
		\centering
		\includegraphics[width=0.5\textwidth]{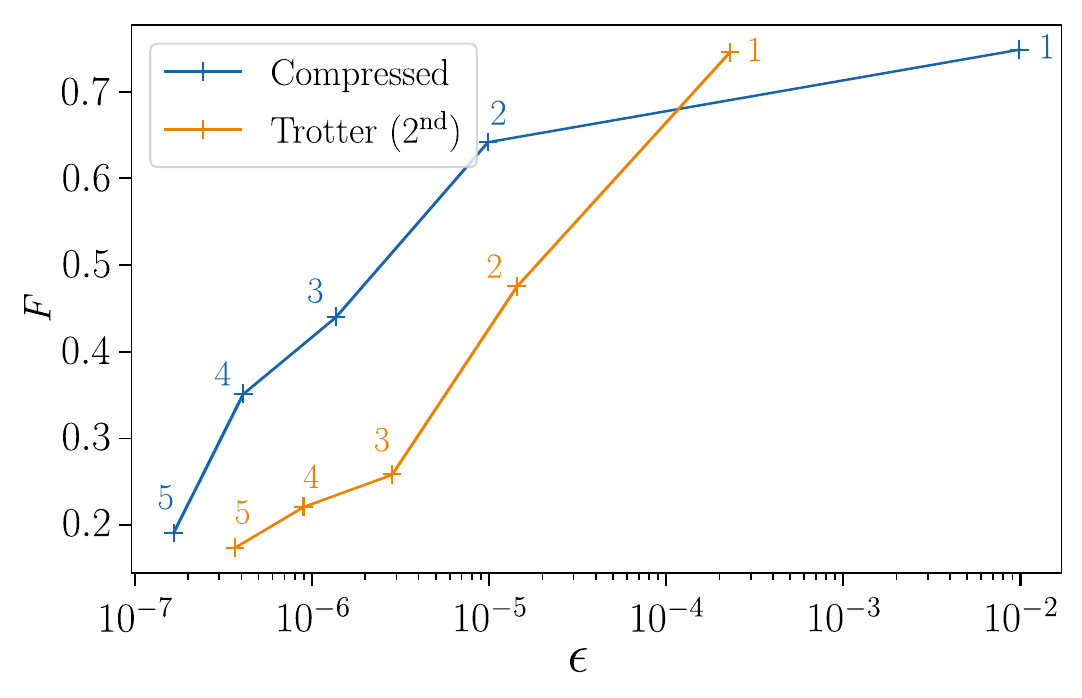}
		\caption{Process fidelity $F$ of the experimentally reconstructed $\chi$ matrices against the approximation error $\epsilon$ of the quantum circuit representation of the unitary operator $U(t=1)$, for both the compressed and Trotter circuits in a system of $L=3$ qubits. The depth of the circuit indicated next to each point is increased to reach a lower approximation error $\epsilon$, and the process fidelity is thus extracted from the corresponding experiments.}
		\label{fig::circ_fid}
	\end{figure}

\emph{3 qubits with open boundary conditions:} We begin by performing full 
quantum process tomography for both circuits in the case of $L=3$ qubits with 
open boundary conditions. The experiments are carried out on the quantum device 
\emph{ibm\_manila}, a 5-qubit chip, where the three qubits are chosen to be 
arranged linearly. Each experiment consists of 1024 shots, and circuit 
transpilation is applied to minimize circuit depth when expressed in the 
device’s native gate set. Fig.~\ref{fig::Chi} shows reconstruction of the 
process matrix $\chi$ for the ideal case (no noise) and from experimental noisy 
measurements of $\vec{d}$.

The left panel displays the theoretical real (top) and imaginary (bottom) parts 
of $\chi(U(t=1))$ for the noiseless case, while the experimental reconstructions 
are shown for both the Trotter and compressed circuits chosen to yield a 
comparable approximation error on the exact evolution operator $U(t=1)$ of 
about $\epsilon \approx 10^{-5}$. These correspond to a 
second-order Trotter decomposition with $n=2$ layers ($dt=0.5$), and a 
compressed circuit also with $n=2$ layers, as confirmed by the matching 
fidelities in Fig.~\ref{fig::Infid}. The reconstructed tomographs capture the 
essential features of the exact process, with the compressed circuit yielding a 
slightly smoother representation due to its reduced number of two-qubit gates. 
We note that the reconstructed processes may not strictly satisfy the 
constraints of completely positive trace preserving maps~\cite{Carteret2008}, as 
noise in $\vec{d}$ can introduce unphysical features such as negative 
eigenvalues in $\chi$. 
While methods exist to enforce physicality, such as variational 
optimization~\cite{Huang_Gao_Jiao_Yan_Zhang_Chen_Zhang_Ji_Jin_2020,Gaikwad_2022}
or alternative forms of process tomography~\cite{White2022}, the comparison of 
the two circuit types presented here remains valid despite small deviations from 
a strictly physical process. Eigenvalues of the superoperator $\Lambda$ with a 
modulus larger than $1$ are unphysical but are useful to estimate the precision 
of the estimate. We note that for $L=3$ qubits, the entire spectrum of $\Lambda$ 
is contained inside the unit disk, but for $L=4$ and SQPT we obtain a few 
eigenvalues with modulus larger than one.

	\begin{figure}[t]
		\centering
		\includegraphics[width=0.5\textwidth]{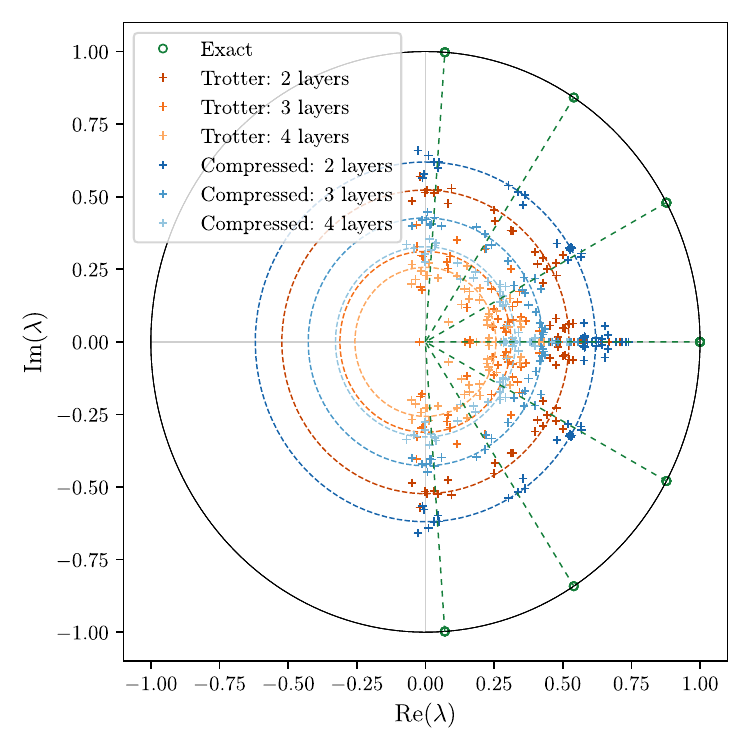}
		\caption{Spectrum of the super-operator $\Lambda$ for $L=3$ and open boundary conditions, where $\lambda$ denote the corresponding eigenvalues. The exact and ideal time-evolution process is unitary and lies on the unit circle of the complex plane. The experimental processes are plotted in red and blue for the Trotter and compressed circuit approximations respectively. Dotted circles show the average modulus of the eigenvalues of an experiment with the corresponding color. The radial green dotted lines are the angles of the clusters of eigenvalues for the ideal quantum process.}
		\label{fig::Process3qubit}
	\end{figure}

	The tomographs provide a three-dimensional visualization of the $4^L$-dimensional process matrix $\chi$, which is difficult to interpret directly. A more global measure of the experimental process quality is the process fidelity~\cite{Gaikwad_2022}, defined as:
	\begin{equation}
		F = \mathrm{Tr}\left[\chi^{\dagger}_{\mathrm{ideal}}\chi^{}_{\mathrm{exp}} \right],
	\end{equation}
	where the process matrices are normalized such that $F \leq 1$, with equality corresponding to perfect agreement. Here, $\chi_{\mathrm{ideal}}$ does not represent the exact time-evolution operator, but its approximation as a quantum circuit. Thus, simulating time evolution on a quantum device involves two distinct sources of error: the approximation of the operator by a finite-depth circuit, and the fidelity with which this circuit is realized on hardware. To compare the two circuit constructions, Fig.~\ref{fig::circ_fid} shows the process fidelity $F$ as a function of the ideal circuit approximation error $\epsilon$. Due to its reduced amount of gates required to achieve a similar accuracy, the compressed circuit performs generally better in terms of fidelity for a comparable approximation. However, beyond three layers for the compressed circuit and two layers for the Trotter circuit, the fidelity drops below $50\%$, rendering the reconstructed process unreliable.

	To gain further insight into how noise affects the experimental process, Fig.~\ref{fig::Process3qubit} displays the spectrum of the super-operator $\Lambda$, with eigenvalues denoted by $\lambda$. In order to reconstruct the process with a reduced contribution of noisy parameters, we consider only the elements of the experimental $\chi$ matrices that are non-zero in the exact ideal one. The first noticeable effect is the presence of non-unitary noise, which manifests as eigenvalues with reduced modulus $|\lambda| < 1$. This deviation from unitarity indicates loss of information and is irreversible, as no physical process can enlarge the modulus of $\lambda$. In this regard, the compressed circuit performs better: for the same number of layers, where we recall that $\epsilon^{}_{\textsc{C}} < \epsilon^{}_{\textsc{T}}$, its eigenvalues remain closer to unity, as highlighted by the guiding dashed circles. Interestingly, for both circuit types—even at relatively low fidelities—the eigenvalues remain well localized in the angular direction. The ideal spectrum features degeneracies arising from system symmetries, and although experimental noise slightly breaks these degeneracies, the reconstructed eigenvalues still cluster around the expected positions. The green dashed radial lines mark the angular locations of these ideal clusters to guide the eye. We observe that the compressed circuits also exhibits lower angular spread than the Trotter ones, suggesting a lower sensitivity to unitary noise sources.

\emph{4 qubits with periodic boundary conditions:} We now consider the Heisenberg chain with periodic boundary conditions for $L=4$ qubits. In this case, full QPT becomes impractical, since the inversion requires handling a $16^4 = 65.536$-dimensional matrix. Instead, we employ more efficient techniques: selective QPT (SQPT)~\cite{Bendersky_Pastawski_Paz_2008, Bendersky_Pastawski_Paz_2009,Schmiegelow_Bendersky_Larotonda_Paz_2011,Perito_Roncaglia_Bendersky_2018,Stefano_Perito_Rebon_2023}, which enables the independent evaluation of specific elements of $\chi$, combined with Pauli twirling~\cite{Wallman_2016, knill2004}, which provides all diagonal elements at reduced cost.

	\begin{figure}[t]
		\centering
		\includegraphics[width=1\columnwidth]{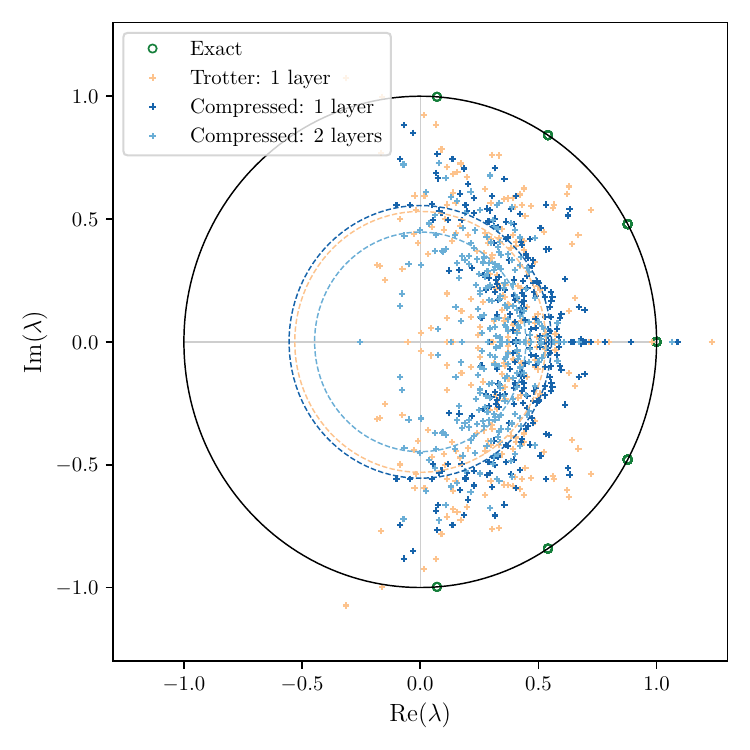}
		\caption{Spectrum of the super-operator $\Lambda$ for $L=4$ and periodic boundary conditions, where $\lambda$ denote the corresponding eigenvalues, plotted as in Fig.~\ref{fig::Process3qubit}. The noisy experimental data leads to the non-physical eigenvalues larger than one.}
		\label{fig::Process4qubit}
	\end{figure}

SQPT offers both advantages and limitations for our purposes. On the one hand, it allows us to target only the symmetry-allowed elements of $\chi$, thereby eliminating spurious non-zero contributions that would otherwise lower the apparent process fidelity in full QPT. On the other hand, SQPT requires the application of additional two-qubit gates on top of the process under study, which increases noise and restricts us to shallower circuits. As a result, process tomography could only be carried out for up to two layers of compressed circuits and a single layer of second-order Trotter circuits, with approximation errors constrained to $\epsilon > 5\times10^{-3}$ (see Fig.~\ref{fig::Infid}).

	\begin{figure}[t]
		\centering
		\includegraphics[width=1\columnwidth]{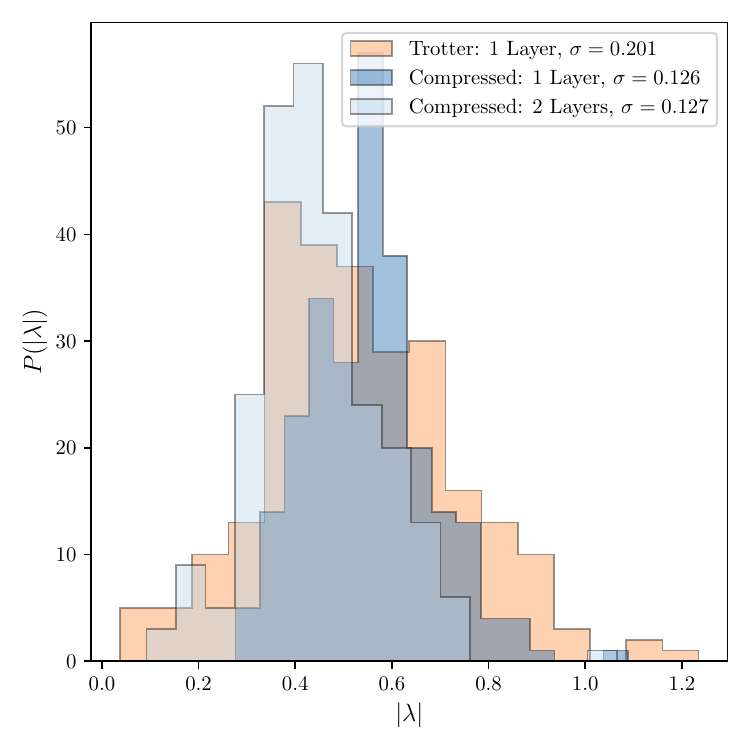}
		\caption{Histogram of the modulus of the eigenvalues of the 4-qubit experimental processes. The Trotter circuit in orange exhibits a wider distribution of the modulus than the compressed ones.}
		\label{fig::Radialdistrib4qubit}
	\end{figure}

The experiments were performed on the 156-qubit \emph{ibm\_fez} device, using 1024 shots. For SQPT, each matrix element requires four experiments (two for the real part and two for the imaginary part; see Appendix~\ref{app:sqpt}). To optimize resources, we partitioned the chip into linear sections of four qubits, reused across the four experiments, so that multiple elements could be measured in parallel. For the compressed circuit, we evaluated the 202 largest non-zero off-diagonal elements (together with their conjugates), while for the Trotter circuit we measured 32 elements; in both cases, the full diagonal was obtained via Pauli twirling. The ideal compressed circuits contains more elements due to the high expressivity of the ansatz, which allows to compress the circuits to shallower forms. The resulting process fidelities are as follows: $F_{\sc c} = 84\%$ for one layer of compressed circuits, $F_{\sc c} = 70\%$ for two layers, and a markedly lower $F_{\sc t} = 57\%$ for one layer of first-order Trotter evolution.

We plot in Fig.~\ref{fig::Process4qubit} the spectrum of the experimentally reconstructed super-operator $\Lambda$, where, as for the three-qubit plot, the average modulus is indicated by dashed colored circles. Compared to the three-qubit case, the eigenvalues here display a broader spread, both radially and axially, as the SQPT circuits have a typically deeper representation. In particular, the Trotter circuits yield a larger fraction of eigenvalues clustered near $|\lambda|=0$, signaling a complete suppression of certain components of the information. Another qualitative difference between the circuit types lies in the occurrence of unphysical eigenvalues with $|\lambda|>1$, which are reduced in the compressed circuits. Because the broader dispersion of eigenvalues makes the mean value less effective at distinguishing the two circuit representations, we additionally show in Fig.~\ref{fig::Radialdistrib4qubit} the distribution of the eigenvalue moduli together with its second moment. This representation highlights more clearly the advantage of the compressed circuits over the deeper Trotter construction: the spectrum of the Trotter time-evolution exhibits a broader distribution, with a second moment nearly twice as large as that of the compressed circuits. Owing to their shallower decomposition and brickwork layout tailored to the hardware, the compressed circuits provide a more robust and noise-resilient realization of the unitary time-evolution operator $U$. While coherent (unitary) noise affects both approaches similarly, depolarizing noise is more pronounced in the Trotter circuits, leading to a greater loss of information during the quantum process.

\section{Conclusion}
\label{Conclusion}

In this work, we experimentally characterized two circuit representations of the time-evolution operator for an $XXX$ spin chain implemented on a superconducting quantum computer. Using both full quantum process tomography and selective process tomography on systems of three and four qubits respectively, we reconstructed the process matrices $\chi$ for different approximation levels of the two circuit types. The approximation error of each circuit representation was controlled by its depth, with deeper circuits achieving higher precision at the cost of reduced process fidelity due to the accumulation of noisy operations. The compressed circuit representation, specifically designed to minimize the approximation error within given depth and hardware connectivity constraints, was hence tested experimentally and demonstrated superior performance on a real noisy device.

For the three-qubit case, the compressed circuits achieved higher process fidelity at comparable approximation error, suggesting that stacking such shallower circuits could enable simulations to reach longer times on quantum hardware. For the four-qubit case, the implementation of SQPT required additional gates, lowering the overall fidelity of both representations. Nonetheless, an analysis of the radial dispersion of the eigenvalues of the super-operator $\Lambda$ revealed that, consistent with the three-qubit results, the compressed circuits were more resilient to depolarizing noise than standard Trotter circuits. These results demonstrate the practical utility of compressed circuit representations in the presence of realistic noise, reinforcing the value of developing and exploring variational ansätze for quantum processes.

\section{Acknowledgments}
\label{Acknowledgments}

We acknowledge support of the Deutsche Forschungsgemeinschaft through the 
cluster of excellence ML4Q (EXC 2004, project-id 390534769), the QuantERA II 
Programme that has received funding from the European Union’s Horizon 2020 
research innovation programme (GA 101017733), the Deutsche 
Forschungsgemeinschaft through the project DQUANT (project-id 499347025), and 
the Deutsche Forschungsgemeinschaft through CRC1639 NuMeriQS (project-id 
511713970) and CRC TR185 OSCAR (project-id 277625399). M. Dinc\u a additionally acknowledges support from the Clarendon Fund and an EPSRC DTP scholarship. We acknowledge the 
use of IBM Quantum Credits for this work. The views expressed are those of the 
authors, and do not reflect the official policy or position of IBM or the IBM
Quantum team.

\bibliography{Bib.bib}

\begin{appendices}
	\section{Selective QPT}
	\label{app:sqpt}

		\begin{figure*}[t]
			\centering
			\includegraphics[width=0.85\textwidth]{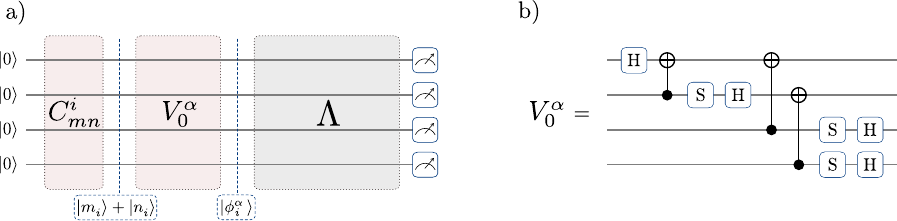}
			\caption{\label{fig::SQPT_circ} a) Circuit overhead required to perform the SQPT of a process $\Lambda$. b) Example of a Clifford circuit implemented to change the basis of a state in a chosen MUB $\alpha$. Two-body \textsc{cnot} gates might be required, thus bringing extra noisy operations to the process $\Lambda$.}
		\end{figure*} 

		The SQPT method relies on the relation between elements of $\chi$ and the average survival probability of specific quantum processes, defined as an integral over the Hilbert space as:
		\begin{equation}
			\begin{split}
				F^{}_{mn} &= \int \mathrm{d}\ket{\phi}\bra{\phi} \Lambda\left(P^{}_{m}|\phi\rangle\langle\phi|P^{}_{n} \right) \ket{\phi}\\
				&= \frac{D\chi^{}_{mn} + \delta^{}_{mn}}{D+1}.
			\end{split}
		\end{equation}
		This formula turns out to be very convenient by transforming the integral into a sum over a set of states forming a $2$-design when the dimension of the Hilbert space $D$ is the power of a prime number \cite{Wooters1989}. Using a set of such states $\{|\phi_{j}\rangle, j=1,\hdots,N\}$, the survival probability reads:
		\begin{equation}
			\label{eq::unphys_proc}
			F^{}_{mn} = \frac{1}{N} \sum\limits^{N}_{j} \langle \phi^{}_{j} | \Lambda\left(\Po{m} |\phi^{}_{j}\rangle\langle\phi^{}_{j}| \Po{n}\right) | \phi^{}_{j}\rangle.
		\end{equation}
		It is clear that the definition above is not practical as the map $\Lambda$ can not be implemented experimentally, since $\Po{m} |\phi^{}_{j}\rangle\langle\phi^{}_{j}| \Po{n}$ is not a valid density matrix. This issue is solved by symmetrizing the previous equation by introducing the following physical maps:
		\begin{equation}
			\label{eq::phys_proc}
			\Lambda\left( \left(\Po{m}+e^{i\gamma}\Po{n}\right)^{\dagger} |\phi^{}_{j}\rangle \langle\phi^{}_{j}| \left(\Po{m}+e^{i\gamma}\Po{n}\right)\right)
		\end{equation}
		where $\gamma \in \{0, \pi, -\pi/2, \pi/2\}$, leading to four distinct fidelities $F^{\gamma}_{mn}$ defined as in Eq.~\eqref{eq::unphys_proc} yet related to a physical map. The desired fidelity is thus obtained by adding them as $F^{}_{mn} = 1/2 \sum_{\gamma} \gamma F^{\gamma}_{mn}$. \\

		It is common to determine the set of states $\{|\phi^{}_{j}\rangle\}$ from the construction of $D+1$ mutually unbiased bases (MUBs) consisting of $D$ states each, which form a $2$-design set by definition. Two orthogonal bases labeled $\alpha$ and $\beta$ are said to be unbiased if the measurement in the first basis doesn't provide any information about measurements in the second one, which can be formulated as $|\langle \phi^{\alpha}_{m} | \phi^{\beta}_{n} \rangle|^{2} = \delta_{\alpha\beta} \delta_{mn} + (1-\delta^{}_{\alpha\beta})1/D$. In the above formula, the label $j$ corresponds now to two indices $(\alpha,m)$.  The $D+1$ MUBs are conveniently obtained by considering a set of $D+1$ commuting Pauli strings stabilizing every states in the corresponding MUB, and finding this appropriate set of Pauli strings amounts to find the stabilizer group for each MUB. In this way, it becomes possible to determine a unitary operation $V^{\alpha}_{0}$ operating the change of basis of a state in the computational basis (labelled by $0$) to the MUB $\alpha$ such that $V^{\alpha}_{0} |\phi^{0}_{m}\rangle = |\phi^{\alpha}_{m}\rangle$. The details for the construction of a given $V^{\alpha}_{0}$ can be found in the appendices of Ref.~\cite{Bendersky_Pastawski_Paz_2009}. 

		In practice, changing the basis of an input state requires the additional implementation of a circuit built with $\mathcal{O} (L^{2})$ gates, which increases the overall depth of the implemented circuit, such that the depth of the circuits implementing the actual operation that we can consider must be reduced to keep an acceptable noise. An example for such a circuit is shown in the right panel of Fig.~\ref{fig::SQPT_circ}. To optimize the implementation of the process in Eq.~\eqref{eq::phys_proc}, we swap the order of the operations:
		\begin{equation}
			\begin{split}
				\left(\Po{m}+e^{i\gamma}\Po{n}\right)^{\dagger} V^{\alpha}_{0}|\phi^{0}_{i}\rangle &= V^{\alpha}_{0} \left(\Tilde{P}^{}_{m}+e^{i\gamma}
				\Tilde{P}_{n}^{}\right)^{\dagger}|\phi^{0}_{i}\rangle  \\
				&= V^{\alpha}_{0} \left(|m^{}_{i}\rangle + |n^{}_{i}\rangle\right) \\
				&= |\phi^{\alpha}_{i}\rangle,
			\end{split}
		\end{equation}
		where $\Tilde{P}_{m} = (V^{\alpha}_{0})^{\dagger} P_{m} V^{\alpha}_{0}$ is a Pauli string, and $|m_{i}\rangle$ and $|n_{i}\rangle$ are states from the computational basis
		in which the relative phase has been absorbed in $|n_{i}\rangle$ for readability. \\

		To summarize the selective method, for each element $\chi_{mn}$, we have to run the four processes $F^{\gamma}_{mn}$. For each process, we first prepare an entangled state in the computational basis $|m^{}_{i}\rangle + |n^{}_{i}\rangle$ through to a circuit $C^{i}_{mn}$ requiring $\mathcal{O}(n)$ gates. Then, we apply the circuit sending states from the computational basis to the MUB $\alpha$, and eventually the physical process that is to be characterized is applied. 

	\section{Pauli twirling}
	\label{Pauli twirls}
		Pauli twirling is a technique that simplifies the description of a quantum process by averaging over random Pauli conjugations~\cite{knill2004,Kern_2005,Geller_2013,Wallman_2016}. For our purpose, we consider the complete set of Pauli strings in order to reconstruct completely the diagonal elements of the process matrix. Concretely, we define the twirled process as:
		\begin{equation}
		\tilde{\Lambda}(\rho_{\mathrm{i}}^{}) = \frac{1}{N}\sum_{m=1}^{N} P_m^{\dagger} \Lambda(P_m^{} \rho_{\mathrm{i}}^{} P^{}_m) P_m^{\dagger},
		\end{equation}
		where the average is taken over all $N=4^L$ Pauli strings. By construction, the resulting map is diagonal in the Pauli basis:
		\begin{equation}
		\tilde{\Lambda}(\rho_{\mathrm{i}}^{}) = \sum_{a=1}^{N} c^{}_a P^{}_a \rho_{\mathrm{i}}^{} P_a^{\dagger},
		\end{equation}
		with coefficients $c_a$ directly related to the diagonal entries of the process matrix $\chi$.
		
		Experimentally, the coefficients $c_a$ can be determined by preparing the input state $\rho_{\mathrm{i}}^{} = P_a$, applying the twirled process, and finally measuring with respect to the same Pauli operator $P_a$:                     
		\begin{equation}
		c_a = \frac{1}{D} \mathrm{Tr}\left[ P_a \tilde{\Lambda}(P_a) \right] = \sum_{m=1}^{N} \chi^{}_{mm} s_{ma}^{},
		\end{equation}                   
		where $s_{ma}=+1$ if $P_a$ commutes with $P_m$ and $s_{ma}=-1$ otherwise.
		
		This relation shows that the diagonal elements of $\chi$ can be collected into a vector $\vec{\chi}_{\mathrm{diag}}$, which is obtained by inverting the known $4^L$-dimensional commutation matrix $s$ and applying it to the vector $\vec{c}$. Importantly, this requires only $4^L$ experimental configurations to recover the entire diagonal of $\chi$, a significant reduction compared to full process tomography. In practice, this approach is combined with SQPT, which is used to reconstruct the off-diagonal elements. An additional advantage is that many Pauli strings communte with each other, meaning that several coefficients $c_a$ can be extracted simultaneously, thereby further lowering the experimental cost. Finally, it also requires almost no gate overhead compared to SQPT, thus improving the quality of the reconstructed process matrix elements on the diagonal compared to the off-diagonal ones.

\end{appendices}

\end{document}